\documentclass[10pt,prb,twocolumn,showpacs]{revtex4}
\usepackage[dvipdfm]{graphicx}
\usepackage{amsmath,amssymb,multirow,natbib}
\newcommand{\be}{\begin{equation}}
\newcommand{\ee}{\end{equation}}

\newcommand{\T}{\mathcal{T}_n}
\newcommand{\Q}{\langle\langle Q_n\rangle\rangle}
\newcommand{\mT}{\langle \mathcal{T}_n\rangle}

\begin{document}
\title{Statistics of quantum transport in chaotic cavities with broken time-reversal symmetry}
\author{Marcel Novaes}
\affiliation{School of Mathematics, University of Bristol, Bristol
BS8 1TW, UK}

\begin{abstract}
The statistical properties of quantum transport through a chaotic cavity are
encoded in the traces $\T={\rm Tr}[(tt^\dag)^n]$, where $t$ is the transmission
matrix. Within the Random Matrix Theory approach, these traces are random
variables whose probability distribution depends on the symmetries of the
system. For the case of broken time-reversal symmetry, we use generalizations
of Selberg's integral and the theory of symmetric polynomials to present
explicit closed expressions for the average value and for the variance of $\T$
for all $n$. In particular, this provides the charge cumulants $\Q$ of all
orders. We also compute the moments $\langle g^n\rangle$ of the conductance
$g=\mathcal{T}_1$. All the results obtained are exact, {\it i.e.} they are
valid for arbitrary numbers of open channels.
\end{abstract}


\maketitle

\section{Introduction}

We consider the problem of electronic transport through a ballistic quantum dot
attached to two ideal leads supporting $N_1$ and $N_2$ open channels. Even at
zero temperature, the electric current as a function of time has a
fundamentally random nature associated with the granularity of charge, and can
naturally be characterized by its statistics \cite{noise}. Its average and
variance, for example, are related respectively to the conductance and the
shot-noise. These quantities have been under experimental investigation for
already quite some time, and when the classical dynamics in the dot is chaotic
they can display characteristic features like weak localization \cite{weak},
universal conductance fluctuations \cite{fluc} and constant Fano factor
\cite{fano}. More recently, much attention has been devoted, both theoretically
\cite{theo} and experimentally \cite{exper}, to higher cumulants comprising the
full counting statistics \cite{prb51hl1995}, which contain more refined
information about the transport process.

This problem can be described, in the Landauer-B\"uttiker scattering approach,
in terms of the unitary scattering matrix $S=\left(\begin{array}{cc}r & t'
\\t & r'
\end{array}\right)$ or in terms of a hermitian matrix $tt^\dag$. The transmission
matrix $t$ has dimension $N_2\times N_1$, where $N_1$ and $N_2$ are the number
of open quantum channels on the entrance and exit lead, respectively. The total
number of channels is $N=N_1+N_2$. The dimensionless conductance is given by
$g={\rm Tr}(tt^\dag)$, and more generally one is interested in the quantities
$\T={\rm Tr}[(tt^\dag)^n]$.

A possible way to model chaotic cavities is to assume $S$ to be drawn at random
from an appropriate ensemble, determined only from the existing symmetries. The
average value of $\T$ over such an ensemble is denoted $\mT$. This is the
random matrix theory approach \cite{rmt}. It treats the eigenvalues of
$tt^\dag$ as correlated random variables, and successfully predicts all the
universal results already mentioned. However, even though experiments may be
done with relatively small channel numbers, so far most explicit expressions
are valid only in the asymptotic limit $N_1,N_2\gg 1$,\cite{rmt,prb75mn2007} or
to the first few terms in a perturbative expansion in $1/N$.

A few particular results exist that are valid for general $N_1,N_2$, like the
average and variance of $g$ \cite{mello,rmt} and the density of transmission
eigenvalues for equal leads $N_1=N_2$. \cite{rmt} More recently, results were
found for: $\langle \mathcal{T}_2\rangle$ and $\langle\mathcal{T}_3\rangle$
\cite{prb73dvs2006,prb75mn2007}; some nonlinear statistics like variance of
shot-noise, skewness and kurtosis of conductance \cite{savin2}; the density of
eigenvalues for unequal leads,\cite{vivo} as well as an expression for $\mT$.
The purpose of this paper is to establish general exact results, of which
several of the above-mentioned ones are particular cases, for systems without
time-reversal symmetry. It is well known that within this symmetry class
leading-order perturbative results sometimes turn out to be exact ({\it e.g.}
for conductance and for the form factor of closed systems). However, this is by
no means usual, and most transport statistics are in general not identical with
their $N\to\infty$ asymptotic limit.

We shall start by computing $\mT$. Our formula is different from the one in
Ref. \onlinecite{vivo}, and computationally superior. We also present the
variance of $\T$ and all moments of the conductance, $\langle
\mathcal{T}_1^n\rangle$. These advancements are made possible by making a
connection with the theory of symmetric polynomials and generalizations of
Selberg's integral.

We must remark that the quantities more readily accessible to measurement are
the charge cumulants $\Q$, which quantify the fluctuations in the amount of
charge transmitted over an interval of time. \cite{prb51hl1995} They are
related to $\mT$ according to the generating function \be \sum_{n=1}^\infty
\frac{x^n}{n!}\Q=-\sum_{m=1}^\infty\frac{(-1)^m}{m}\langle \mathcal{T}_m\rangle
(e^x-1)^m.\ee Our exact expression for $\mT$ therefore provides cumulants $\Q$
of arbitrary order.

For broken time-reveral systems, random matrix theory predicts that the
non-zero eigenvalues $\{T_1,...T_{N_1}\}$ of $tt^\dag$ behave statistically
like random numbers distributed between $0$ and $1$ according to the joint
probability distribution \cite{rmt,for2} \be\label{dist}
\mathcal{P}(T)=\mathcal{N}|\Delta(T)|^2 \prod_jT_j^{\alpha},\ee where
$\Delta(T)=\prod_{i<j}(T_i-T_j)$ is the Vandermonde determinant,
$\alpha=N_2-N_1$ (we assume $N_1\le N_2$) measures the asymmetry between the
leads and $\mathcal{N}$ is a normalization constant. Averages are therefore
obtained as \be \langle f(T)\rangle =\int_{\mathcal{C}}
\mathcal{P}(T)f(T)dT,\ee where $\mathcal{C}$ is the hypercube $[0,1]^{\otimes
N_1}$. By integrating out all but one of the variables one obtains the
eigenvalue density $\rho(T)$. An exact expression for $\rho(T)$ can be written
down using Jacobi polynomials, \cite{vivo,rho} \be
\rho(T)=T^{\alpha}\sum_{j=0}^{N_1-1}(\alpha+2j+1)[P_j^{(\alpha,0)}(1-2T)]^2,\ee
and hence $\mT=N_1\int_0^1 T^n\rho(T)dT$. Our result (\ref{mom}) may be seen as
an explicit solution to this integral.

\section{Generalizations of the Selberg integral}

We start by fixing some notation (a basic reference is
[\onlinecite{macdonald}]). A non-increasing sequence of positive integers
$\lambda=(\lambda_1,\lambda_2,\ldots)$ is said to be a partition of $n$ if
$\sum_i\lambda_i=n$. This is indicated by $\lambda \vdash n$. The number of
parts in $\lambda$ is called its length and denoted by $\ell(\lambda)$. The set
of all partitions of $n$ is $P(n)$. With every partition $\lambda$ we can
associate a monomial $x^\lambda=x_1^{\lambda_1}x_2^{\lambda_2}\cdots$. The
symmetric polynomial in $k$ variables $m^{(k)}_\lambda(x)$ is the sum of all
distinct monomials obtainable from $x^\lambda$ by permutation of the $x$'s. For
example, take $n=k=2$. Then $ m_{(2)}(x)=x_1^2+x_2^2$ and $
m_{(1,1)}(x)=x_1x_2.$ The set of functions $\{m^{(k)}_\lambda(x),\lambda \in
P(n)\}$ forms a basis for the vector space of homogeneous symmetric polynomials
of degree $n$ in $k$ variables.

Another basis for this space is composed by Schur functions
$s_\lambda^{(k)}(x)$. If $\lambda \vdash n\le k$, \be
s_\lambda^{(k)}(x)=\frac{1}{\Delta(x)}{\rm
det}\left(x_i^{k+\lambda_j-j}\right)_{1\le i,j\le k}.\ee Let us use again
$n=k=2$ as an example. We then have $ s_{(2)}(x)=x_1x_2$ and $
s_{(1,1)}(x)=x_1^2+x_2^2+x_1x_2.$ These two basis are of course related by a
linear transformation, $s_\lambda^{(k)}=\sum_{\mu\vdash
n}K_{\lambda,\mu}m_{\mu}^{(k)}.$ The matrices $K$ are called Kostka matrices.

Let us from now on take $k=N_1$, {\it i.e.} consider only polynomials in $N_1$
variables, and no longer write any superscript. Let \be
[x]_\lambda=\prod_{i=1}^{\ell(\lambda)}\frac{(x+\lambda_i-i)!}{(x-i)!}\ee be a
generalization of the usual rising factorial \be (x)_a=x(x+1)\cdots(x+a-1).\ee
Our favorite examples are now $[x]_{(2)}=x(x+1)$ and $[x]_{(1,1)}=x(x-1)$. Let
us also define the function \be
H_\lambda=\frac{\prod_i(n+\lambda_i-i)!}{\prod_{i<j}(\lambda_i-\lambda_j+j-i)}.\ee
This is equal to the product of all hook lengths of the partition
$\lambda$,\cite{macdonald} but that is not relevant here. Kaneko \cite{kaneko}
and Kadell \cite{kadell} have proved a generalization of Selberg's integral
which implies that \be\label{kadell} \langle
s_\lambda\rangle=\int_{\mathcal{C}} s_\lambda(T)\mathcal{P}(T)
dT=\frac{[N_1]_{\lambda}[N_2]_{\lambda}}{[N]_{\lambda}H_{\lambda}}.\ee Other
generalizations are known. For example, Hua \cite{hua} (see also
[\onlinecite{kadell2}]) has shown that \be\label{hua} \int_{\mathcal{C}}
s_\lambda(T)s_\mu(T)\mathcal{P}(T) dT= \langle s_\lambda\rangle \langle
s_\mu\rangle\{N+1\}_{\lambda,\mu},\ee
where\be\{N\}_{\lambda,\mu}=\prod_{i,j}\frac{(N-i-j+\lambda_i)(N-i-j+\mu_j)}
{(N-i-j+\lambda_i+\mu_j)(N-i-j)}.\ee

\section{Results}
\subsection{Average of $\T$}

Making use of (\ref{kadell}) one can in principle obtain the average value of
any symmetric polynomial of the transmission eigenvalues. For example, it is
known \cite{macdonald} that \be\label{T2m}
\T=m_{(n)}(T)=\sum_{p=0}^{n-1}(-1)^ps_{(n-p,1^p)}(T), \ee where the notation
$1^p$ means that the number $1$ appears $p$ times. Substituting into
(\ref{kadell}) gives the remarkably simple result for the counting statistics
\be\label{mom} \mT=\sum_{p=0}^{n-1}G_{n,p},\ee where \be
G_{n,p}=\frac{(-1)^p}{n!}{n-1 \choose p}\frac{(N_1-p)_n(N_2-p)_n}{(N-p)_n}.\ee
The sum (\ref{mom}) allows in principle the calculation of any linear
statistic. It has the merit of having only $n$ terms, in contrast with the
result in [\onlinecite{vivo}] where the number of terms in the sum grows with
$N_1,N_2$. In Fig.1 (top) we plot $\mT/N$ for a few values of $n$, as a
function of $N_1$ for $N_2=20$. All curves have a maximum around $N_1\approx
N_2$, where they are approximately equal to ${2n\choose n}/4^n$.

Let us denote by $\mT_\infty$ the asymptotic form of $\mT$ when both channel
numbers are large, $N_1,N_2\gg 1$. These functions are given by
\cite{prb75mn2007} \be\label{asymp}
\mT_\infty=N\xi\sum_{m=0}^{n-1}\frac{(-1)^m}{m+1}{2m \choose m}{n-1 \choose m}
\xi^m,\ee where we have defined the finite variable $\xi=N_1N_2/N^2$.
Interestingly, a derivation of this expression directly from (\ref{mom}) does
not seem to be trivial. Notice that $\mT_\infty={2n\choose n}/4^n$ for
$N_1=N_2$. In Fig.1 (bottom) we compare the exact result (\ref{mom}) with the
asymptotics (\ref{asymp}) for $N_2=1$, in which case $\langle
\T\rangle=N_1/(N_1+n)$. We can see that these quantities may differ
significantly if both channel numbers are close to unity, specially for higher
values of $n$.

\begin{figure}[t]
\centerline{\includegraphics[clip,scale=1.25]{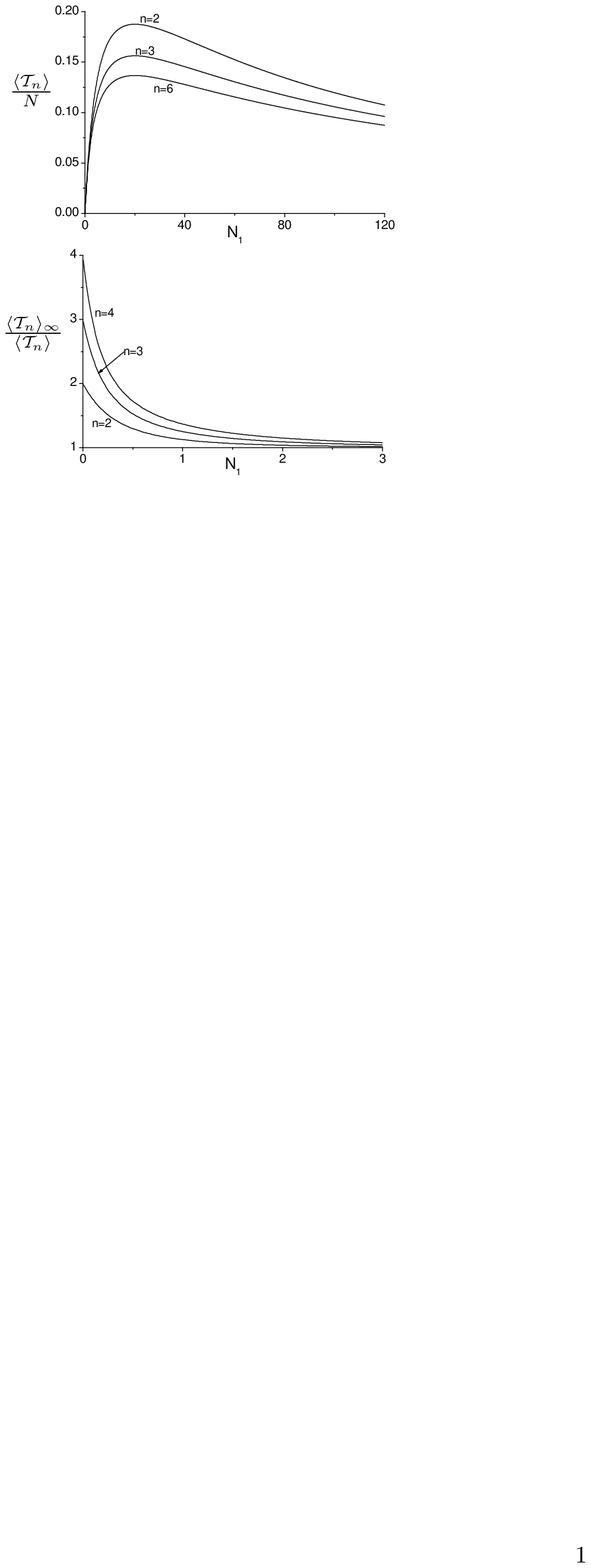}}
\caption{Top: some of the counting statistics for $N_2=20$. The value at the maximum $N_1\approx N_2$ is close to
${2n\choose n}/4^n$. Bottom: Comparison between exact result (\ref{mom})
and asymptotic limit (\ref{asymp}) for $N_2=1$. }
\end{figure}

\begin{figure}[t] \centerline{\includegraphics[clip,scale=0.9]{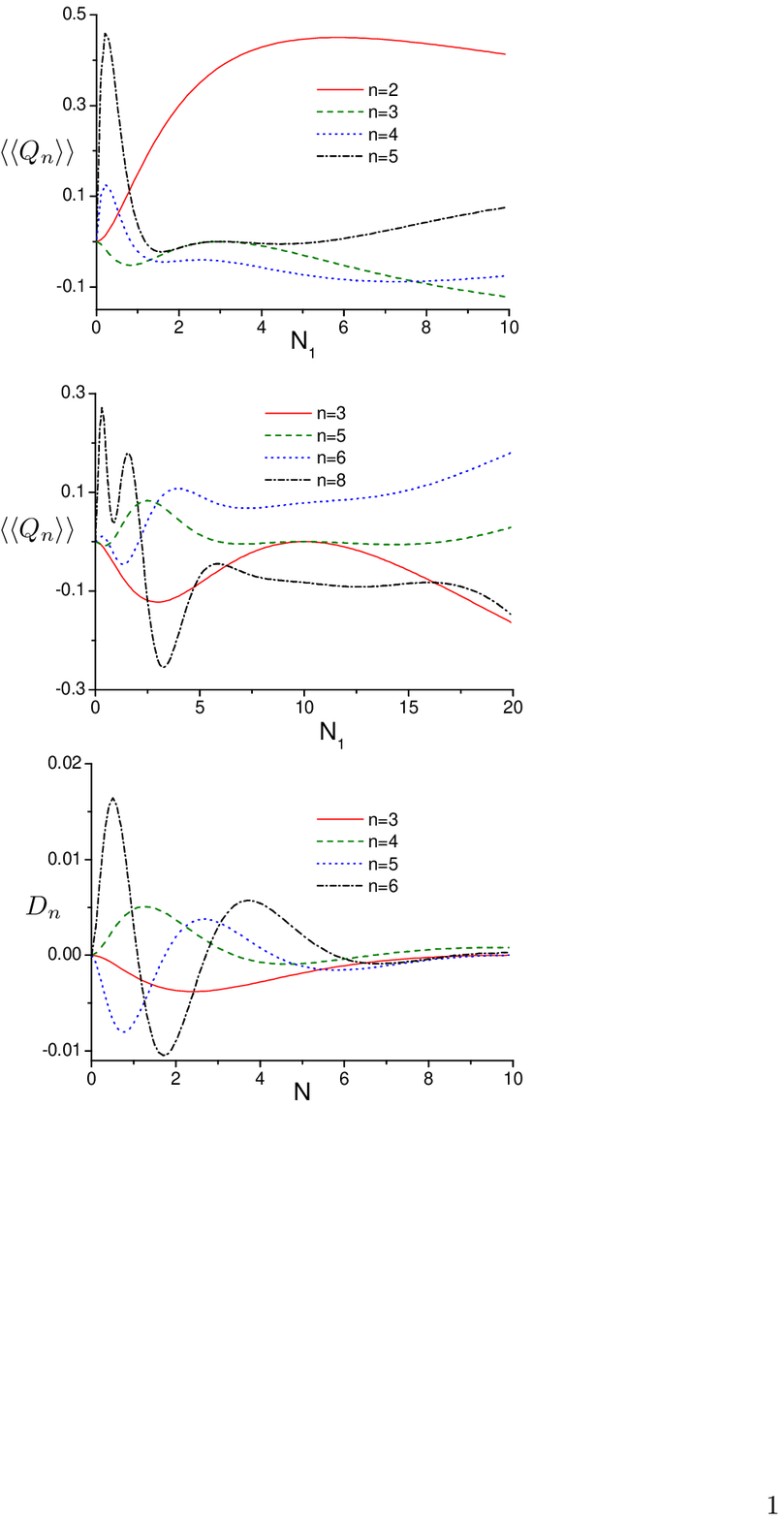}}
\caption{(color online) Charge cumulants $\Q$
as functions of $N_1$ for $N_2=3$ (top) and $N_2=10$ (middle). In the bottom
panel we plot the difference $D_n=\Q-\Q_\infty$ between the exact result and its
asymptotic expression for equal leads.}
\end{figure}

We can obtain from (\ref{mom}) the set of charge cumulants $\Q$, which are
given by \be\label{charge}
\Q=\sum_{m=1}^{n}C_{n,m}\langle\mathcal{T}_{m}\rangle, \ee with the
coefficients \be C_{n,m}=\sum_{j=0}^{m-1}(-1)^j{m-1\choose j}(j+1)^{n-1}.\ee
Figure 2 shows some plots of $\Q$. The odd ones are always zero for equal
leads, $N_1=N_2$, and contain a factor $(N_2-N_1)^2$ in general. Except for the
second one (shot-noise), all $\Q$ oscillate and even change sign as functions
of $N_1$ for fixed $N_2$. The asymptotic form of these quantities for large
channel numbers, which we denote $\Q_\infty$, can be derived by using
$\mT_\infty$ instead of $\mT$ in (\ref{charge}). That yields
\cite{prb75mn2007}\be
\Q_\infty=-N\xi\sum_{m=0}^{n-1}(-1)^{n+m}\frac{(2m)!}{(m-1)!}S_{n-1,m}\xi^m,\ee
where $S_{n,m}$ are the Stirling numbers of the second kind. In the last panel
of Fig.2 we compare the exact and asymptotic results for $N_1=N_2=N$. In
contrast to $\mT$, for which the asymptotic value is always above the exact
one, here we see that the difference $\Q-\Q_\infty$ oscillates as a function of
$N$, with a number of zeros that increases with $n$.

\subsection{Variance of $\T$}

We now turn to the calculation of ${\rm var}(\T)$. First, we note that,
analogously to (\ref{T2m}), we can write \be\label{var}
\T^2=\sum_{p,q=0}^{n-1}(-1)^{p+q}s_{(n-p,1^p)}(T)s_{(n-q,1^q)}(T).\ee Using
(\ref{hua}) we see that ${\rm var}(\T)=\langle\T^2\rangle-\langle\T\rangle^2$
is given by \be {\rm var}(\T)=\sum_{p,q=0}^{n-1}
\frac{n^2G_{n,p}G_{n,q}}{(N+2n-p-q-1)(N-p-q-1)}.\ee This generalizes to $n>1$
the well known result for conductance fluctuations,
var$(\mathcal{T}_1)=N_1^2N_2^2/N^2(N^2-1)$. Of course, we could just as easily
have computed the value of $\langle \T\mathcal{T}_m\rangle$ for any $n,m$.

In Fig.3 we plot ${\rm var}(\T)$ as a function of $N_1$ for the case $N_2=20$,
for a few values of $n$. The function increases mildly with $n$, and is
monotonic in $N_1$. They all saturate at a finite value as $N_1\to\infty$. For
$N_1=N_2$ the many-channels limit can be obtained explicitly by other means
\cite{pier} and is given by \be\label{varsa} {\rm
var}(\T)_\infty=\frac{(2n-1)\Gamma(n+1/2)\Gamma(n-1/2)}{8\pi n
(\Gamma(n))^2}.\ee This quantity starts at the value $1/16$ (variance of the
conductance), and saturates at $1/4\pi$ for large $n$.

\begin{figure}[t]
\centerline{\includegraphics[clip,scale=1.,angle=90]{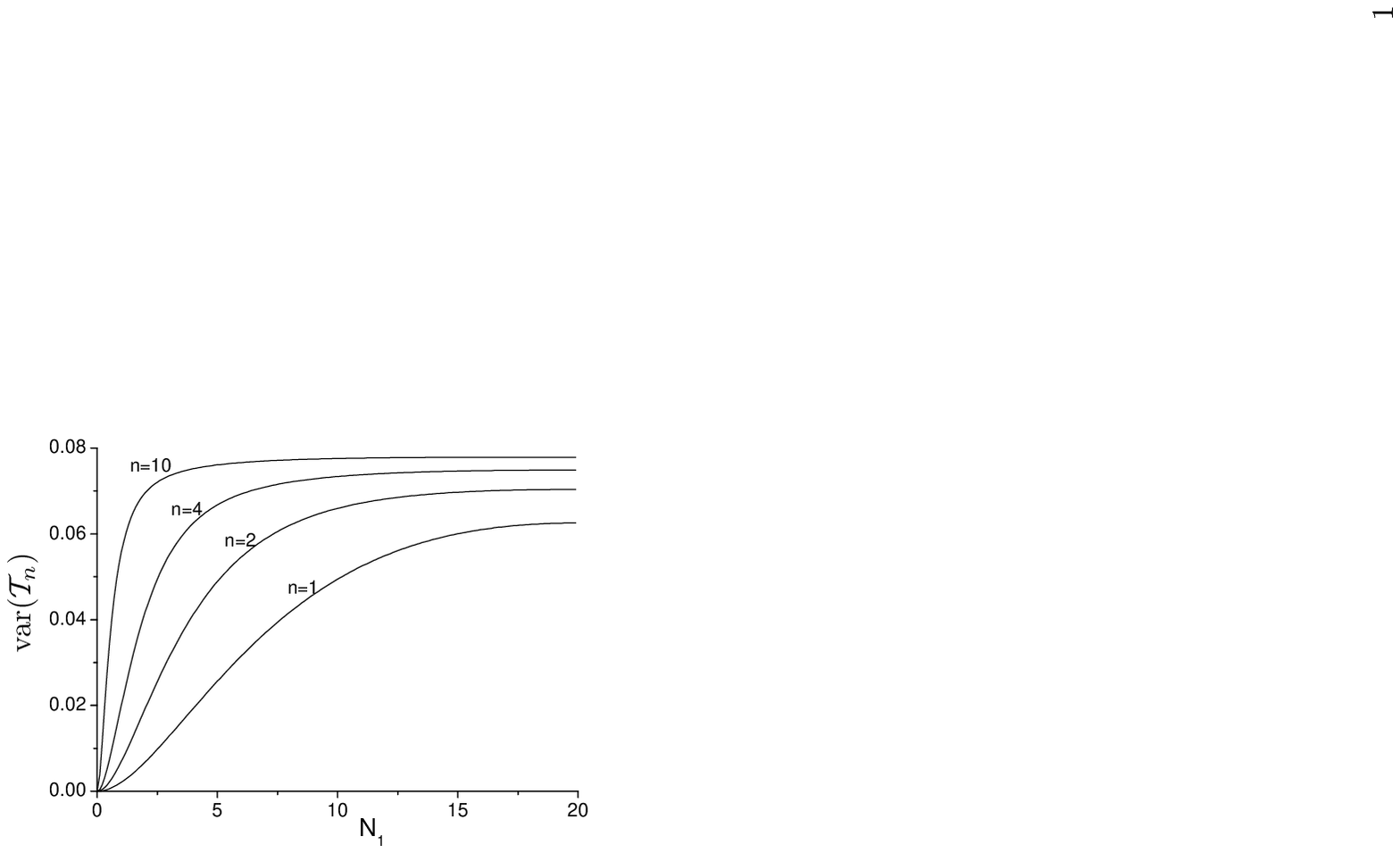}} \caption{Variances var($\T$) as functions of $N_1$ for $N_2=20$.}
\end{figure}

\begin{figure}[!t]
\centerline{\includegraphics[clip,scale=0.95,angle=90]{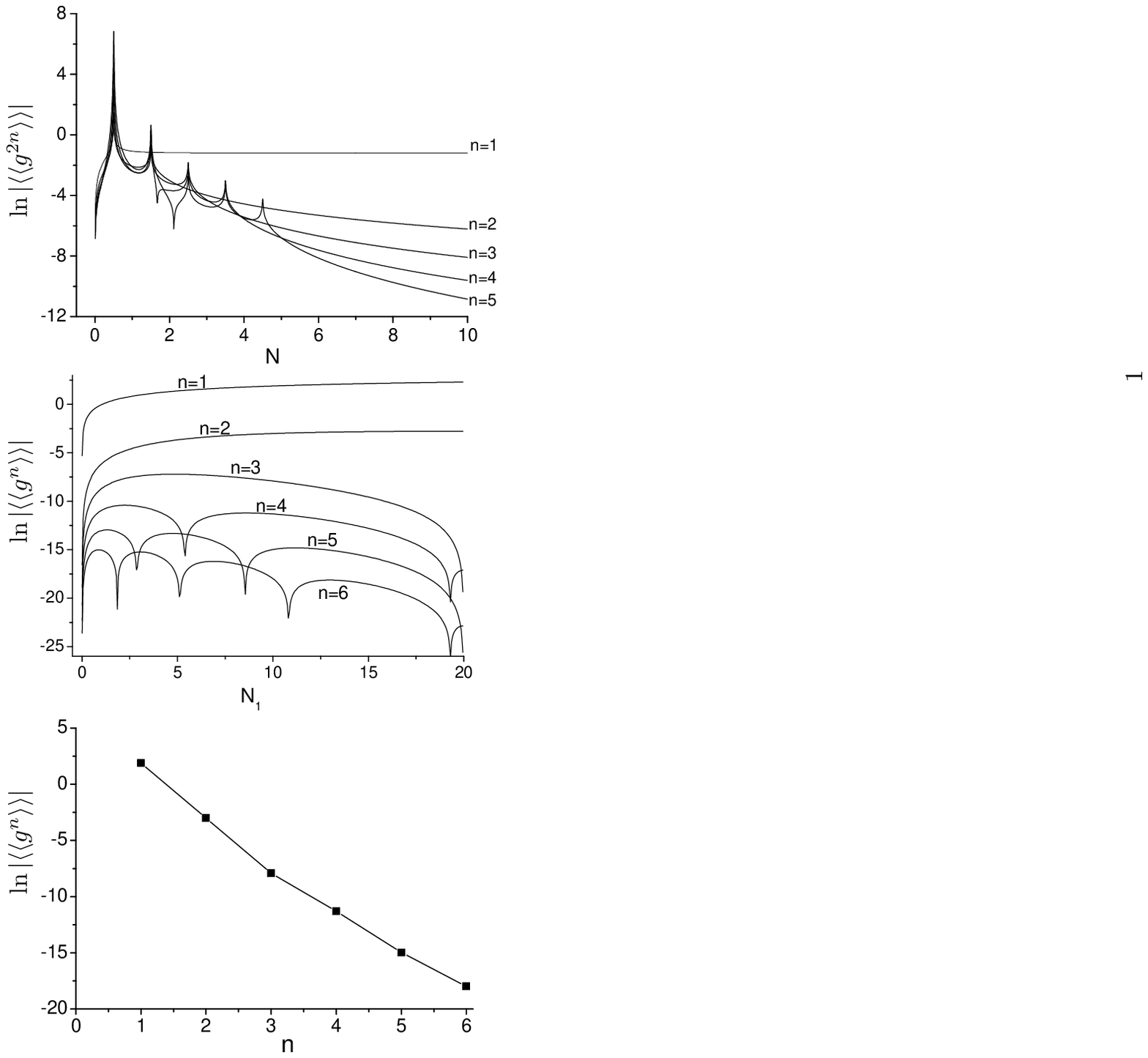}}
\caption{Top: Even cumulants $\langle \langle
g^{2n}\rangle \rangle$ of the conductance for equal leads $N_1=N_2=N$
(odd cumulants are identically zero). Middle: $\langle
\langle g^{n}\rangle \rangle$ as a function of $N_1$ for $N_2=20$. Bottom: Decay of
$\langle\langle g^{n}\rangle \rangle$ with $n$ for $N_2=2N_1=20$. Notice the logarithmic scale in all cases.}
\end{figure}

\subsection{Moments of the conductance}

It is known that for large channel numbers the probability distribution of the
conductance $g= \mathcal{T}_1$ approaches a Gaussian, while for small $N_1,N_2$
there are deviations. The last quantity we shall compute is the set of all
moments $\langle g^n\rangle$ characterizing this distribution. Forrester
\cite{forres} has proved (see also [\onlinecite{yan}]) that
\be\label{forres}\left\langle\prod_j(1-xT_j)^{-a}\right\rangle=\sum_\lambda
\frac{[a]_\lambda [N_2]_\lambda}{[N]_\lambda H_\lambda}s_\lambda(x),\ee where
we have an infinite sum over partitions. Let us perform the change of variables
$x=y/a$ and take the limit $a\to\infty$. It is well known \cite{macdonald}
that, if $\lambda\vdash n$, then \be [a]_\lambda\approx a^n, \quad
s_\lambda(y/a)\approx
\left(\frac{y}{a}\right)^n\frac{[N_1]_\lambda}{H_\lambda}, \quad a\to\infty.\ee
Using $\lim_{a\to\infty}(1-yT/a)^{-a}=e^{yT}$ on the left hand side and the
simplifications above on the right hand side, equation (\ref{forres}) provides
a generating function for the quantities we are looking for. The result is
\be\label{cond} \langle g^n\rangle=n!\sum_{\lambda\vdash n}
\frac{[N_1]_\lambda[N_2]_\lambda}{[N]_\lambda H_\lambda^2}.\ee

From these moments one may compute the corresponding cumulants $\langle\langle
g^n \rangle\rangle$. For a Gaussian distribution, $\langle\langle g^n
\rangle\rangle=0$ for $n>2$. From the symmetry of the distribution
(\ref{dist}), all odd cumulants with $n>1$ vanish identically if $N_1=N_2$ (in
general, they contain the factor $(N_1-N_2)^2$). Figure 4 (top) presents some
of the even cumulants $\langle\langle g^{2n} \rangle\rangle$ for $N_1=N_2=N$.
We see that $\langle\langle g^2 \rangle\rangle$ saturates at the expected value
$1/16$. Although the others seem to have somewhat singular behaviour for $N<n$,
only integer values of $N$ are actually physical. They all decrease rapidly
after $N>n$, and already for $N=6$ the distribution function of the conductance
is Gaussian to a good approximation. In Fig.4 (middle) we see the first few
cumulants as functions of $N_1$ for $N_2=20$, while Fig.4 (bottom) shows how
fast the value of $\langle \langle g^{n}\rangle \rangle$ decays with $n$, for
$N_2=2N_1=20$.

\section{Conclusions}

We have extended the random matrix theory of quantum transport in ballistic
chaotic cavities connected to two ideal leads, by obtaining exact results for
several statistical quantities. All our formulas as expressed in simple closed
form. Even though we consider systems with broken time-reversal (TR) symmetry,
we have seen that exact results are in general different from their large-N
asymptotic limit.

Unfortunately, the calculations presented here are not immediately
generalizable to TR symmetric systems, for which we expect the difference
between exact and asymptotic results to be more important. This is because the
role played here by Schur functions is played in that case by zonal
polynomials, and much less is known about the analogues of Kostka matrices.
This certainly deserves further investigation (actually, relation
(\ref{forres}) has been proven\cite{forres} for other symmetry classes, so the
moments of the conductance can in fact be obtained in those cases). Also, it
would be interesting to obtain systematic exact results for more general
nonlinear statistics.

After this work was completed, another paper appeared \cite{kanzieper} in which
the probability distribution of conductance is studied for broken TR. In
particular, they obtain an efficient recurrence relation for the cumulants of
$g$. It would be important to clarify what are the relations between those
results and the present work.

\section*{Acknowledgments}

I thank P. Vivo for telling me about (\ref{varsa}) before publication, and J.P.
Keating for pointing out [\onlinecite{kaneko}] and [\onlinecite{yan}]. Both of
them, and D.V. Savin, are also thanked for stimulating discussions. This work
was supported by EPSRC.


\begin{thebibliography}{99}

\bibitem{noise} {\it Quantum Noise in Mesoscopic Physics}, edited by Yu.V.
    Nazarov (Dordrecht, Kluwer, 2003).

\bibitem{weak} A.M. Chang, H.U. Baranger, L.N. Pfeiffer and K.W. West, Phys.
    Rev. Lett. {\bf 73}, 2111 (1994).

\bibitem{fluc} C.M. Marcus, A.J. Rimberg, R.M. Westervelt, P.F. Hopkins, and
    A.C. Gossard, Phys. Rev. Lett. {\bf 69}, 506 (1992).

\bibitem{fano} S. Oberholzer, E.V. Sukhorukov, C. Strunk, C. Sch\"onenberger,
    T. Heinzel and M. Holland, Phys. Rev. Lett. {\bf 86}, 2114 (2001).

\bibitem{theo} Ya.M. Blanter, H. Schomerus and C.W.J. Beenakker, Physica E {\bf
    11}, 1 (2001); Yu.V. Nazarov and D.A. Bagrets, Phys. Rev. Lett. {\bf 88},
    196801 (2002); S. Pilgram, A.N. Jordan, E.V. Sukhorukov and M. B\"uttiker, {\it ibid}
     {\bf 90}, 206801 (2003); E.V. Sukhorukov and O.M. Bulashenko, {\it ibid} {\bf 94},
116803 (2005); S. Pilgram, P. Samuelsson, H. F\"orster and M. B\"uttiker {\it
ibid} {\bf 97}, 066801 (2006); O.M. Bulashenko, J. Stat. Mech. P08013 (2005).

\bibitem{exper} W. Lu, Z. Ji, L. Pfeiffer, K.W. West and A.J. Rimberg,
    Nature {\bf 423}, 422 (2003); J. Bylander, T. Duty and P. Delsing, {\it ibid}
{\bf 434}, 361 (2005); T. Fujisawa, T. Hayashi, Y. Hirayama, H.D. Cheong and Y.
H. Jeong, Appl. Phys. Lett. {\bf 84}, 2343 (2004); R. Schleser, E. Ruh, T. Ihn,
K. Ensslin, D.C. Driscoll and A.C. Gossard, {\it ibid} {\bf 85}, 2005 (2004);
E.V. Sukhorukov, A.N. Jordan, S. Gustavsson, R. Leturcq, T. Ihn and K. Ensslin,
Nature Phys. {\bf 3}, 243 (2007).

\bibitem{prb51hl1995} L.S. Levitov, and G.B. Lesovik, JETP Lett. {\bf 58}, 230
    (1993); H. Lee, L.S. Levitov and A.Yu. Yakovets, Phys. Rev. B {\bf 51},
    4079 (1995).

\bibitem{rmt} C.W.J. Beenakker, Phys. Rev. Lett. {\bf 70}, 1155 (1993); H.U.
    Baranger and P.A. Mello, {\it ibid} {\bf 73}, 142 (1994); R.A. Jalabert,
    J.-L. Pichard and C.W.J. Beenakker, Europhys. Lett. {\bf 27}, 255 (1994);
    P.W. Brouwer, and C.W.J. Beenakker, J. Math. Phys. {\bf
    37}, 4904 (1996); C.W.J. Beenakker, Rev. Mod. Phys. {\bf 69}, 731 (1997).

\bibitem{prb75mn2007} M. Novaes, Phys. Rev. B {\bf 75}, 073304 (2007).

\bibitem{mello} P.A. Mello, J. Phys. A: Math. Gen. {\bf 23}, 4061 (1990).

\bibitem{prb73dvs2006} D.V. Savin and H.-J. Sommers, Phys. Rev. B {\bf 73},
    081307(R) (2006).

\bibitem{savin2} H.-J. Sommers, W. Wieczorek and D.V. Savin, Acta Phys.
    Pol. A {\bf 112}, 691 (2007); D.V. Savin, H.-J. Sommers and W. Wieczorek,
Phys. Rev. B {\bf 77}, 125332 (2008).

\bibitem{vivo} P. Vivo and E. Vivo, J. Phys. A {\bf 41}, 122004 (2008).

\bibitem{for2} P.J. Forrester, J. Phys. A {\bf 39}, 6861 (2006).

\bibitem{rho} J.E.F. Ara\'ujo and A.M.S Mac\^edo, Phys. Rev. B {\bf 58},
    R13379 (1998).

\bibitem{macdonald} {\it Symmetric functions and Hall polynomials}, I.G.
    MacDonald (Oxford, Oxford University Press, 1998).

\bibitem{kaneko} J. Kaneko, SIAM J. Math. Anal. {\bf 24}, 1086 (1993).

\bibitem{kadell} K.W.J. Kadell, Adv. Math. {\bf 130}, 33 (1997).

\bibitem{hua} {\it Harmonic analysis of functions of several complex
    variables in the classical domains}, L.K. Hua (Providence, Amer. Math. Soc., 1963).

\bibitem{kadell2} K.W.J. Kadell, Compos. Math. {\bf 87}, 5 (1993).

\bibitem{pier} P. Vivo, S.N. Majumdar and O. Bohigas, unpublished.

\bibitem{forres} P.J. Forrester, Nucl. Phys. B {\bf 416}, 377
    (1994).

\bibitem{yan} Z. Yan, Canad. J. Math. {\bf 44}, 1317 (1992).

\bibitem{kanzieper}  V. Al. Osipov and E. Kanzieper, arXiv:0806.2784 (2008).

\end{thebibliography}
\end{document}